\documentclass[a4paper,11pt]{article}
\usepackage{jheppub} 
\usepackage{lineno}

\title{\boldmath Unified equation for massless spin fields and new definitions of key spin coefficients}

\author{Zhong-Heng Li}
\affiliation{Black Hole and Gravitational Wave Group,\\
 Zhejiang University of Technology,\\
 No. 288 Liuhe Road, Hangzhou 310023, China}

\emailAdd{zhli@zjut.edu.cn}

\abstract{Whether studying gravitational waves from extreme mass ratio inspirals or exploring the analogy between massless spin-particle waves, black hole perturbation theory proves indispensable. At the heart of developing a universal perturbation framework for such problems lies the challenge of formulating a coordinate-independent, unified wave equation that is universally applicable to any black hole spacetime.
This paper resolves this central issue in type-D spacetimes by introducing a generating function $H$ and establishing new definitions for the key spin coefficients. Specifically, the spin coefficients $\rho$, $\mu$, $\tau$, and $\pi$ are redefined, respectively, as the directional derivatives of the logarithm of the generating function along the null tetrad ($l^{\mu}$, $n^{\mu}$, $m^{\mu}$, $\bar{m}^{\mu}$), and the field quantities are rescaled using $H$. It is thereby found that the field equations governing massless particles of spins $0$, $1/2$, $1$, $3/2$, and $2$ in arbitrary type-D black hole spacetimes can all be described by a single, unified equation. This finding is particularly remarkable, as unifying these field equations is already a significant challenge in flat spacetime, let alone in the intricate spacetime around black holes.
Consequently, this work will inevitably prompt a re-examination of the shared characteristics among various types of particles in black hole spacetimes. Meanwhile, we verify the correctness of the new definition for the spin coefficients, and provide the explicit form of the unified equation for nearly all known type-D black hole backgrounds. This lays a solid foundation not only for studying gravitational waves from extreme mass ratio inspirals but also for exploring the analogy between massless spin-particle waves in any type-D black hole background.}

\begin{document}


\maketitle
\flushbottom

\section{Introduction}
\label{sec:intro}

Since humanity's first detection of gravitational waves on September 14, 2015 [1], the field of gravitational wave astronomy has advanced rapidly. To date, the LIGO Scientific Collaboration has confirmed over 100 gravitational wave events originating from binary black hole mergers (with a few exceptions). For decades, the Teukolsky master equation [2, 3] has served as a fundamental tool for modeling gravitational waveforms produced during black hole mergers [4]. However, A commonly occurring inspiral scenario in nature is where one black hole is much lighter than the other, forming what is called an ``extreme mass ratio inspiral (EMRI)''. EMRIs are a key source of gravitational waves for future space-based detectors like LISA [5, 6], Taiji [7], and TianQin [8]. EMRI signals enable ultra-precise measurements of central object parameters [9], allowing stringent tests of black hole models -- ultimately verifying whether the object is a Kerr black hole or a non-Kerr alternative.

The exciting prospect of observing EMRIs with LISA, Taiji, and TianQin has, over the past 25 years, driven concerted theoretical efforts to develop numerous models for EMRIs. These models span studies of astrophysical environments [10-18] to tests of General Relativity and beyond [19-34]. Despite extensive discussions, a fundamental equation for studying various types of black holes is still lacking. Therefore, it is necessary to establish a gravitational wave master equation with broader applicability. The first objective of this paper is to solve this problem. The second objective is to establish an equation for studying the analogy between massless particle waves.

Analogy is one of the basic thinking methods in the process of understanding objective things, its physical basis is that different physical systems obey the same dynamical evolution equations. Similarity analyses provide cross-fertilization of ideas among different branches of science. The development of gravitational and electromagnetic theories serves as a model example in this regard.

Since both Newton's universal law of gravitation and Coulomb's law are the inverse square of the distance, the study of the analogy between gravity and electromagnetism has a long history. In 1849, Faraday [35] designed a series of experiments, similar to those for induction of electricity by magnetism, to detect
what he termed ``gravelectric current'' in a helix of wire. In 1865, Maxwell [36] attempted to develop a vector theory of gravity by exploring the possibility of formulating the gravitational theory in a manner analogous to the equations of electromagnetism. In the 1870s, Hozm$\ddot{u}$ller [37] and Tisserand [38, 39] used the so-called gravitational magnetic field to explain the precession of Mercury's perihelion.

In 1915, Einstein established the general theory of relativity, making it difficult to imagine any similarities between the geometric theory of gravity and electromagnetism. Surprisingly, in 1953, Matte [40] derived a Maxwell-like structure for the linearized general theory of relativity.  After persistent efforts over a long period, a specialized theory known as gravitoelectromagnetism [41, 42] was established in the weak field approximation. Thus, two important questions naturally arise: First, do the perturbation equations of gravity and electromagnetism have the same form in a strong gravitational background? Second, can the analogies between gravitational and electromagnetic fields be extended to other massless spin fields? These questions have seen considerable progress in research over the past few decades.

The Teukolsky master equation, derived using the Newman-Penrose formalism [43], achieves two fundamental results: it first decouples perturbations in the Kerr metric and then unifies the massless scalar field, Weyl neutrino field, electromagnetic field, and gravitational field within a single framework. This equation not only serves as an indispensable tool for investigating phenomena such as Kerr black hole formation through binary mergers but also offers profound insights into the characteristics of strong-field gravity. These contributions mark Teukolsky's work as a landmark achievement in theoretical astrophysics. Subsequent studies have extended the Teukolsky formalism to other black hole spacetimes [44-47]. Inspired by these advances, we conjecture that the wave equations governing all massless spin particles can be unified into a single statement in specific Petrov-type black hole spacetimes. Verification of this conjecture would simultaneously fulfill the dual objectives of this work. To verify this conjecture, we recall Chandrasekhar's observation that ``It is a remarkable fact that the black-hole solutions of general relativity are all of Petrov type D'' [48].
In this study, we therefore establish a unified description for all massless fields with spin $s\leq2$ in Petrov type-D spacetimes. For consistency, black holes in such spacetimes are hereafter termed type-D black holes. To achieve this goal, we must first find new representations of certain spin coefficients.

This paper is organized as follows. In section 2, we provide a brief overview of the aspects of the Newman-Penrose formalism relevant to our research.
In section 3, we introduce a generating function $H$ and propose new definitions and interpretations for the spin coefficients $\rho$, $\mu$, $\tau$, and $\pi$. By employing these reformulated spin coefficients and rescaled field quantities using $H$, we establish a unified equation that simultaneously describes the massless scalar, Weyl neutrino, electromagnetic, Rarita-Schwinger, and gravitational fields in Petrov type-D spacetimes.
In section 4, we verify the correctness of the new definitions for the spin coefficients, and derive the explicit form of the unified equation for nearly all known type-D black hole backgrounds. We conclude the paper in section 5. In appendix A, we give the spin coefficients for general spherically symmetric spacetimes, general Vaidya-type spacetimes, and the Pleba$\acute{n}$ski-Demia$\acute{n}$ski metric, the complete family of black hole-like spacetimes, the Kerr-Newman-de Sitter spacetime, and the variable-mass Kerr metric. In appendix B, we present three tables that demonstrate the partial black hole solutions contained within general spherically symmetric spacetimes, general Vaidya-type spacetimes, and the complete family of black hole-like spacetimes, respectively. Appendix C provides the Bianchi identities in the vacuum case.

Throughout this paper, we adopt the following conventions. The complex conjugate of a quantity $z$ is denoted by $\bar{z}$, and the spacetime metric has signature (+, -, -, -). The Newman-Penrose null tetrad is given by ($l^{\mu}$, $n^{\mu}$, $m^{\mu}$, $\bar{m}^{\mu}$), normalized such that $l_{\mu} n^{\mu} = 1$, $m_{\mu} \bar{m}^{\mu} = -1$, with all other inner products vanishing. The spin coefficients are denoted by $\kappa$, $\lambda$, $\sigma$, $\nu$, $\rho$, $\tau$, $\mu$, $\pi$, $\alpha$, $\beta$, $\gamma$, $\varepsilon$; the Weyl scalars by $\psi_{0}$, $\psi_{1}$, $\psi_{2}$, $\psi_{3}$, $\psi_{4}$; and the directional derivatives by $D$, $\Delta$, $\delta$, $\bar{\delta}$. The spin quantum number is $s$ and the spin weight is $p$. Field quantities are written as $\chi_{p}^{(s)}$ and source terms as $T_{p}^{(s)}$, with their rescaled versions denoted by $\Phi_{p}$ and $T_{p}$, respectively.
The covariant derivative is denoted by $\nabla_{\mu}$, the generating function is $H$, and the spin-coefficient connection is represented by $L^{\mu}$.

\section{Null tetrad, spin coefficients and Weyl scalars}
\label{sec:intro}
To ensure rigorous generality in our approach, we employ the Newman-Penrose formalism [43], which is a tetrad formalism based on a set of four null vectors. Within this formalism, twelve complex spin coefficients, five complex scalar functions encoding the Weyl tensor, three complex Maxwell scalars, and nine functions encoding the tracefree Ricci tensor are introduced. Additionally, the Ricci scalar is replaced by a real scalar $\Lambda=R/24$. However, to avoid confusion, we will not adopt this substitution in the subsequent equations. In this section, rather than providing an exhaustive review, we briefly summarize some results pertinent to our work.

The tetrad consists of two real null vectors, $l_{\mu}$ and $n_{\mu}$, and a pair of complex null vectors, $m_{\mu}$ and $\bar{m}_{\mu}$, which satisfy the orthonormal conditions,
\begin{eqnarray}\label{2.1}
&&l_{\mu}l^{\mu}=n_{\mu}n^{\mu}=m_{\mu}m^{\mu}=\bar{m}_{\mu}\bar{m}^{\mu}=0, \nonumber\\
&&l_{\mu}n^{\mu}=-m_{\mu}\bar{m}^{\mu}=1, \nonumber\\
&&l_{\mu}m^{\mu}=l_{\mu}\bar{m}^{\mu}=n_{\mu}m^{\mu}=n_{\mu}\bar{m}^{\mu}=0.
\end{eqnarray}
The indexes are raised and lowered using the global metric $g_{\mu\nu}$, which can be expressed in terms of null vectors as follows:
\begin{eqnarray}\label{2.2}
g_{\mu\nu}=2l_{(\mu}n_{\nu)}-2m_{(\mu}\bar{m}_{\nu)}.
\end{eqnarray}
Also, the metric can be written more compactly:
\begin{eqnarray}\label{2.3}
g^{\mu\nu}=\eta^{ij}\lambda^{\mu}\,_{i}\lambda^{\nu}\,_{j}.
\end{eqnarray}
Here, the tetrad index $i$ is raised and lowered using the flat metric $\eta_{ij}$, and $\lambda^{\mu}\,_{i}$ is defined by
\begin{eqnarray}\label{2.4}
\lambda^{\mu}\,_{i}=(l^{\mu}, n^{\mu}, m^{\mu}, \bar{m}^{\mu}),
\end{eqnarray}
with
\begin{equation}\label{2.5}
\begin{array}{ll}
\eta^{ij}=\left(\begin{array}{lll}
0\,\,\,\,\,\,1\,\,\,\,\,\,0\,\,\,\,\,\,0\\
1\,\,\,\,\,\,0\,\,\,\,\,\,0\,\,\,\,\,\,0\\
0\,\,\,\,\,\,0\,\,\,\,\,\,0\,\,-1\\
0\,\,\,\,\,\,0\,\,-1\,\,\,\,\,\,0
\end{array}
\right).
\end{array}
\end{equation}
Note that the tetrad admits certain freedoms, such as

(a) Null rotation around $l^{\mu}$:
\begin{eqnarray}\label{2.6}
&&l^{\mu} \to l^{\mu}, \nonumber\\
&&m^{\mu} \to m^{\mu} + \zeta l^{\mu}, \nonumber\\
&&n^{\mu} \to n^{\mu} + \bar{\zeta} m^{\mu} + \zeta \bar{m}^{\mu} + \zeta \bar{\zeta} l^{\mu};
\end{eqnarray}

(b) null rotation around $n^{\mu}$:
\begin{eqnarray}\label{2.7}
&&n^{\mu} \to n^{\mu}, \nonumber\\
&&m^{\mu} \to m^{\mu} + \xi n^{\mu}, \nonumber\\
&&l^{\mu} \to l^{\mu} + \bar{\xi} m^{\mu} + \xi \bar{m}^{\mu} + \xi \bar{\xi} n^{\mu};
\end{eqnarray}

(c) boost in the $l^{\mu}-n^{\mu}$ plane and spatial rotation in the $m^{\mu}-\bar{m}^{\mu}$ plane:
\begin{eqnarray}\label{2.8}
&&l^{\mu} \to \textsf{A} l^{\mu}, \nonumber\\
&&n^{\mu} \to \textsf{A}^{-1} n^{\mu}, \nonumber\\
&&m^\mu \to e^{i\theta} m^\mu.
\end{eqnarray}
Here $\zeta$, and $\xi$ are complex numbers, and $\textsf{A}$ and $\theta$ are real.

In the Newman-Penrose formalism, the twelve spin coefficients are defined by the following expressions:
\begin{eqnarray}\label{2.9}
&&\kappa=\bigtriangledown_{\nu}l_{\mu}m^{\mu}l^{\nu}, \quad \lambda=-\bigtriangledown_{\nu}n_{\mu}\bar{m}^{\mu}\bar{m}^{\nu},\nonumber\\
&&\sigma=\bigtriangledown_{\nu}l_{\mu}m^{\mu}m^{\nu}, \quad \nu=-\bigtriangledown_{\nu}n_{\mu}\bar{m}^{\mu}n^{\nu},\nonumber\\
&&\rho=\bigtriangledown_{\nu}l_{\mu}m^{\mu}\bar{m}^{\nu}, \quad \tau=\bigtriangledown_{\nu}l_{\mu}m^{\mu}n^{\nu},\nonumber\\
&&\mu=-\bigtriangledown_{\nu}n_{\mu}\bar{m}^{\mu}m^{\nu}, \quad \pi=-\bigtriangledown_{\nu}n_{\mu}\bar{m}^{\mu}l^{\nu},\nonumber\\
&&\alpha=\frac{1}{2}(\bigtriangledown_{\nu}l_{\mu}n^{\mu}\bar{m}^{\nu}-\bigtriangledown_{\nu}m_{\mu}\bar{m}^{\mu}\bar{m}^{\nu}),\nonumber\\
&&\beta=\frac{1}{2}(\bigtriangledown_{\nu}l_{\mu}n^{\mu}m^{\nu}-\bigtriangledown_{\nu}m_{\mu}\bar{m}^{\mu}m^{\nu}),\nonumber\\
&&\gamma=\frac{1}{2}(\bigtriangledown_{\nu}l_{\mu}n^{\mu}n^{\nu}-\bigtriangledown_{\nu}m_{\mu}\bar{m}^{\mu}n^{\nu}),\nonumber\\
&&\varepsilon=\frac{1}{2}(\bigtriangledown_{\nu}l_{\mu}n^{\mu}l^{\nu}-\bigtriangledown_{\nu}m_{\mu}\bar{m}^{\mu}l^{\nu}),
\end{eqnarray}
where $\nabla_{\nu}$ denotes the covariant derivative. Many of spin coefficients have direct geometric significance [49]. For instance, the vanishing of $\kappa$ is the condition for the integral curves of $l^{\mu}$ to be geodesic, while, if $\sigma$ is also zero, this congruence of geodesics is shear free. The same role is played by $\nu$ and $\lambda$ for the $n^{\mu}$-congruence.

In the Newman-Penrose formalism, the ten independent components of the Weyl tensor are completely determined by the five complex Weyl scalars, which are defined as follows:
\begin{eqnarray}\label{2.10}
&&\psi_{0}=-C_{\mu\nu\rho\sigma}l^{\mu}m^{\nu}l^{\rho}m^{\sigma},\nonumber\\
&&\psi_{1}=-C_{\mu\nu\rho\sigma}l^{\mu}n^{\nu}l^{\rho}m^{\sigma},\nonumber\\
&&\psi_{2}=-\frac{1}{2}C_{\mu\nu\rho\sigma}(l^{\mu}n^{\nu}l^{\rho}n^{\sigma}-l^{\mu}n^{\nu}m^{\rho}\bar{m}^{\sigma}),\nonumber\\
&&\psi_{3}=-C_{\mu\nu\rho\sigma}\bar{m}^{\mu}n^{\nu}l^{\rho}n^{\sigma},\nonumber\\
&&\psi_{4}=-C_{\mu\nu\rho\sigma}\bar{m}^{\mu}n^{\nu}\bar{m}^{\rho}n^{\sigma},
\end{eqnarray}
where $C_{\mu\nu\rho\sigma}$ is the Weyl tensor, which satisfies
\begin{equation}\label{2.11}
\begin{aligned}
 R_{\mu\nu\rho\sigma}=C_{\mu\nu\rho\sigma}+\frac{1}{2}(g_{\mu\rho}R_{\nu\sigma}-g_{\mu\sigma}R_{\nu\rho}-g_{\nu\rho}R_{\mu\sigma}
 +g_{\nu\sigma}R_{\mu\rho})+\frac{1}{6}(g_{\mu\sigma}g_{\nu\rho}-g_{\mu\rho}g_{\nu\sigma})R.
\end{aligned}
\end{equation}
Here $R_{\mu\nu\rho\sigma}$, $R_{\mu\nu}$, and $R$ are the Riemann tensor, Ricci tensor, and scalar curvature, respectively.

Each of the Weyl scalars carries a specific physical interpretation as follows [50]. $\psi_{0}$ is a transverse component propagating in the $n^{\mu}$ direction. $\psi_{1}$ is a longitudinal component in the $n^{\mu}$ direction. $\psi_{2}$ is a Coulomb-like component. $\psi_{3}$ is a longitudinal component in the $l^{\mu}$ direction. $\psi_{4}$ is a transverse component propagating in the $l^{\mu}$ direction.

Petrov type D is characterized by the existence of two double principal null directions, $n^{\mu}$ and $l^{\mu}$, thus having the following formulas for the Weyl scalars and the spin coefficients [51, 52]:
\begin{eqnarray}\label{2.12}
\psi_{0}=\psi_{1}=\psi_{3}=\psi_{4}=0,
\end{eqnarray}
\begin{eqnarray}\label{2.13}
\kappa=\sigma=\lambda=\nu=0.
\end{eqnarray}

For vacuum spacetimes, if either Eq. (2.12) or Eq. (2.13) holds, the other necessarily holds as well. Because if the spin coefficients, $\kappa, \sigma, \lambda$ and $\nu$ vanish, we can conclude on the basis of the Goldberg-sachs theorem [53] that the Weyl scalars, $\psi_0, \psi_1, \psi_3 ,\text{and}\ \psi_4 $ must vanish in the chosen basis. The Weyl scalar $\psi_2$ does not, however, vanish. The perturbation quantities $\psi_{0}^{B}$ and $\psi_{4}^{B}$ of $\psi_{0}$ and $\psi_{4}$ are invariant under gauge transformations and infinitesimal tetrad rotations [3], and are therefore completely measurable physical quantities, which represent the gravitational waves of spin weight 2 and spin weight -2, respectively.

\section{Unified equation based on new spin coefficient definitions}
\label{sec:intro}

In this section, we focus on generating functions, new definitions of the spin coefficients, and the unified description of all massless fields with spin $s\leq2$ in Petrov type-D spacetimes.

\subsection{New definitions of spin coefficients $\rho$, $\mu$, $\tau$ and $\pi$}

In 1963, Newman and Penrose introduced 12 complex spin coefficients represented by Eq. (2.9). Among these, the real and imaginary parts of $\rho$ are, respectively (minus), the expansion and the twist of the congruence of integral curves of $l^{\mu}$; and $|\sigma|$  is a measure of the degree of the shear [49].
For over half a century, the definitions of the Newman-Penrose spin coefficients and their associated geometric interpretations have been widely used, with no alternative definitions or explanations proposed. Herein, we present an entirely new definition for the spin coefficients $\rho$, $\mu$, $\tau$ and $\pi$, expressed as
\begin{eqnarray}\label{3.1}
\rho=-D\ln H, \quad \mu=\Delta\ln H, \quad \tau=-\delta\ln H, \quad \pi=\bar{\delta}\ln H.
\end{eqnarray}
Here $D, \Delta,$ and $\delta$ are the directional derivatives defined by
\begin{equation}\label{3.2}
D=l^{\mu}\partial_{\mu}, \quad \Delta=n^{\mu}\partial_{\mu}, \quad \delta=m^{\mu}\partial_{\mu}, \quad \bar{\delta}=\bar{m}^{\mu}\partial_{\mu};
\end{equation}
$H$ is a complex function which compactly specifies the spin coefficients $\rho$, $\mu$, $\tau$ and $\pi$ through partial derivatives of its logarithm. Furthermore, as we will demonstrate, $H$ determines a transformation relation of wave functions, leading to its designation as the generating function in this paper.

In type-D vacuum spacetimes (with or without a cosmological constant), the validity of Eq. (3.1) can be straightforwardly proven. By employing Eqs. (2.12) and (2.13), the Bianchi identity (C.1) reduces to: $D\psi_{2}=3\rho\psi_{2}$, $-\bar{\delta}\psi_{2}=3\pi\psi_{2}$, $\delta\psi_{2}=3\tau\psi_{2}$, $\Delta\psi_{2}=-3\mu\psi_{2}$. Without loss of generality, we may assume $\psi_{2}=const\cdot H^{-3}$. Under this assumption, the simplified Bianchi identities exactly reproduce Eq. (3.1).

It should be noted that in type D non-vacuum spacetimes, the relation $\psi_{2}=const\cdot H^{-3}$ generally fails to hold; nevertheless, the definition provided in Eq. (3.1) remains valid. For example, in subsections 4.1 and 4.2, we demonstrate -- using different coordinate systems and null tetrads -- that definition (3.1) holds in any spherically symmetric spacetime, regardless of the matter content (e.g., vacuum, electromagnetic fields, pure radiation, dust, perfect fluids, or other forms). In subsections 4.3 and 4.4, we further prove that (3.1) remains valid in the Pleba$\acute{n}$ski-Demia$\acute{n}$ski family of spacetimes under various coordinate and tetrad choices. Notably, this class includes all type-D vacuum and aligned non-null Einstein-Maxwell solutions with a possible cosmological constant. Additionally, in section 5, we provide examples showing that (3.1) continues to hold in non-type D spacetimes.

Equation (3.1) admits the following physical interpretations: $\rho$ is the rate of decrease of the logarithm of the generating function in the $l^{\mu}$ direction, $\mu$ is the rate of increase of the logarithm of the generating function in the $n^{\mu}$ direction, $\tau$ is the rate of decrease of the logarithm of the generating function in the $m^{\mu}$ direction, $\pi$ is the rate of increase of the logarithm of the generating function in the $\bar{m}^{\mu}$ direction.

An inverse problem involves determining how to obtain a generating function from the null tetrad and spin coefficients. Clearly, the equations we need to solve constitute a first-order differential system:
\begin{equation}\label{3.3}
\begin{array}{ll}
\left(\begin{array}{lll}
l^{0}\,\,\,\,\,\,\,\,\,\,\,l^{1}\,\,\,\,\,\,\,\,\,l^{2}\,\,\,\,\,\,\,\,\,l^{3}\\
n^{0}\,\,\,\,\,\,\,\,n^{1}\,\,\,\,\,\,\,\,n^{2}\,\,\,\,\,\,\,\,n^{3}\\
m^{0}\,\,\,\,\,\,m^{1}\,\,\,\,\,\,m^{2}\,\,\,\,\,\,m^{3}\\
\bar{m}^{0}\,\,\,\,\,\,\bar{m}^{1}\,\,\,\,\,\,\bar{m}^{2}\,\,\,\,\,\,\bar{m}^{3}
\end{array}
\right)
\left(\begin{array}{lll}
\partial_{0}\ln H\\
\partial_{1}\ln H\\
\partial_{2}\ln H\\
\partial_{3}\ln H
\end{array}
\right)=\left(\begin{array}{lll}
-\rho\\
\,\,\,\mu\\
-\tau\\
\,\,\,\pi
\end{array}
\right).
\end{array}
\end{equation}

\subsection{Decoupled equations}

In general, the components of a spin field in curved spacetimes are coupled. However, for the Weyl neutrino, electromagnetic, Rarita-Schwinger, and gravitational fields in a Petrov type-D spacetime, the equations for each field can be simplified into two decoupled equations in the case of perturbations, as outlined below.

Decoupled Weyl neutrino equations ($s=1/2$) [3]  are given by
\begin{eqnarray}\label{3.4}
&&[(D+\bar{\varepsilon}-\rho-\bar{\rho})(\Delta-\gamma+\mu)
-(\delta-\bar{\alpha}-\tau+\bar{\pi})(\bar{\delta}-\alpha+\pi)]\chi_{0}=0,\nonumber\\
&&[(\Delta-\bar{\gamma}+\mu+\bar{\mu})(D+\varepsilon-\rho)
-(\bar{\delta}+\bar{\beta}+\pi-\bar{\tau})(\delta+\beta-\tau)]\chi_{1}=0.
\end{eqnarray}

Decoupled electromagnetic equations ($s=1$) [3] are given by
\begin{eqnarray}\label{3.5}
&&[(D-\varepsilon+\bar{\varepsilon}-2\rho-\bar{\rho})(\Delta-2\gamma+\mu)
-(\delta+\bar{\pi}-\bar{\alpha}-\beta-2\tau)(\bar{\delta}+\pi-2\alpha)]\phi_{0}=2\pi J_{0},\nonumber\\
&&[(\Delta+\gamma-\bar{\gamma}+2\mu+\bar{\mu})(D+2\varepsilon-\rho)
-(\bar{\delta}-\bar{\tau}+\bar{\beta}+\alpha+2\pi)(\delta-\tau+2\beta)\phi_{2}=2\pi J_{2}.\nonumber\\
\end{eqnarray}

Decoupled Rarita-Schwinger equations ($s=3/2$) [54] are given by
\begin{eqnarray}\label{3.6}
&&[(D-2\varepsilon+\bar{\varepsilon}-3\rho-\bar{\rho})(\Delta-3\gamma+\mu)
-(\delta+\bar{\pi}-\bar{\alpha}-2\beta-3\tau)(\bar{\delta}+\pi-3\alpha)
-\psi_2]H_{000}=0,\nonumber\\
&&[(\Delta+2\gamma-\bar{\gamma}+3\mu+\bar{\mu})(D+3\varepsilon-\rho)
-(\bar{\delta}-\bar{\tau}+\bar{\beta}+2\alpha+3\pi)(\delta-\tau+3\beta)
-\psi_2]H_{111}=0.\nonumber\\
\end{eqnarray}

Decoupled gravitational equations ($s=2$) [3] are given by
\begin{eqnarray}\label{3.7}
&&[(D-3\varepsilon+\bar{\varepsilon}-4\rho-\bar{\rho})(\Delta-4\gamma+\mu)-(\delta+\bar{\pi}-\bar{\alpha}-3\beta-4\tau)(\bar{\delta}+\pi-4\alpha)-3\psi_2]\psi_{0}^{B}
=4\pi T_{0},\nonumber\\
&&[(\Delta+3\gamma-\bar{\gamma}+4\mu+\bar{\mu})(D+4\varepsilon-\rho)
-(\bar{\delta}-\bar{\tau}+\bar{\beta}+3\alpha+4\pi)(\delta-\tau+4\beta)
-3\psi_2]\psi_{4}^{B}=4\pi T_{4}.\nonumber\\
\end{eqnarray}

Here $J_{0}$, $J_{4}$, $T_{0}$ and $T_{4}$ are the source terms; and the ``$\pi$'' on the right-hand side of Eqs. (3.5) and (3.7) is the constant Pi. We use $p$ to represent the spin weight. In each pair of equations from (3.4) to (3.7), the first equation is for the spin weight of $p=s$, while the other one is for $p=-s$.

\subsection{A unified equation}

In reviewing all analogical studies, the central challenge is to develop a unified framework that can describe the dynamical equations of all analogous systems with a single statement.

To derive an equation that uniformly describes Eqs. (3.4)-(3.7), we standardize the notation: wave functions, source terms, and constants are represented by $\chi_{p}^{(s)}$, $T_{p}^{(s)}$, and $\kappa_{s}$, respectively, then we have $\chi_{1/2}^{(1/2)}=\chi_{0}$, $\chi_{-1/2}^{(1/2)}=\chi_{1}$, $T_{1/2}^{(1/2)}=0$, $T_{-1/2}^{(1/2)}=0$, $\kappa_{1/2}=0$; $\chi_{1}^{(1)}=\phi_{0}$, $\chi_{-1}^{(1)}=\phi_{2}$, $T_{1}^{(1)}=J_{0}$, $T_{-1}^{(1)}=J_{2}$, $\kappa_{1}=2\pi$, etc. These notations allow us to combine all equations in (3.4)-(3.7) with spin weight $p = s$ into
\begin{eqnarray}\label{3.8}
&\big\{[D-(2s-1)\varepsilon+\bar{\varepsilon}-2s\rho-\bar{\rho}](\Delta-2s\gamma+\mu) \nonumber\\
&-[\delta+\bar{\pi}-\bar{\alpha}-(2s-1)\beta-2s\tau](\bar{\delta}+\pi-2s\alpha) \nonumber\\
&-(2s-1)(s-1)\psi_{2}\big\}\chi^{(s)}_{s}=\kappa_{s}T^{(s)}_{s},
\end{eqnarray}
and those with spin weight $p =-s$ into
\begin{eqnarray}\label{3.9}
&\big\{[\Delta+(2s-1)\gamma-\bar{\gamma}+2s\mu+\bar{\mu}](D+2s\varepsilon-\rho) \nonumber\\
&-[\bar{\delta}-\bar{\tau}+\bar{\beta}+(2s-1)\alpha+2s\pi](\delta-\tau+2s\beta) \nonumber\\
&-(2s-1)(s-1)\psi_{2}\big\}\chi^{(s)}_{-s}=\kappa_{s}T^{(s)}_{-s}.
\end{eqnarray}
When source terms vanish, Eqs. (3.8) and (3.9) simplify to Eqs. (6) and (7) of Ref. [55], respectively. It should be noted that the source terms are essential for investigating gravitational waves from EMRIs. Additionally, Eq. (3.9) is particularly significant as it determines the Weyl scalar $\psi_{4}$. In the EMRI problem, this scalar enables the extraction of key physical quantities: the radiative fluxes (both at infinity and at the horizon), as well as the two gravitational wave polarizations $h_{+}$ and $h_{\times}$.

Although Eqs. (3.8) and (3.9) are straightforward to obtain, combining them into a single unified expression proves exceptionally challenging and necessitates the use of the newly defined spin coefficients (Eq. (3.1)). Below, we undertake the task of merging Eqs. (3.8) and (3.9).

Using $H$, determined by Eq. (3.1), we introduce the rescaled field quantities $\Phi_{p}$, defined by
\begin{eqnarray}\label{3.10}
\chi_{p}^{(s)}=H^{p-s}\Phi_{p}.
\end{eqnarray}

As early as 1972, Teukolsky [2] introduced the rescaled wave functions $\rho^{-4} \psi_{4}$ and $\rho^{-2} \phi_{2}$ (where the expression for $\rho$ is given in (A.6)) to obtain separable equations for the gravitational wave $\psi_{4}$ and electromagnetic wave $\phi_{2}$ in Kerr spacetime. With the null tetrad adopted by Teukolsky [2], the generating function is $H = -\rho^{-1}$. According to Eq. (3.10), $\Phi_{-2} = H^{4} \chi_{-2}^{(2)} = \rho^{-4} \psi_{4}$, \quad $\Phi_{-1} = H^{2} \chi_{-1}^{(1)} = \rho^{-2} \phi_{2}$,
which exactly coincide with the rescaled wave functions introduced by Teukolsky. Thus, Teukolsky's rescaling of $\psi_{4}$ and $\phi_{2}$ is merely a special case of Eq. (3.10). Incidentally, substituting $H = -\rho^{-1}$ into Eq. (3.1) yields spin coefficients $\rho$, $\mu$, $\tau$, and $\pi$ that are fully consistent with those in Ref. [2].

In 1978, Chandrasekhar [56] (see also Ref. [48], p. 432, Eq. (6)) introduced rescalings of $\psi_{0}$, $\psi_{1}$, $\psi_{3}$, and $\psi_{4}$ in Kerr spacetime to render the equations for $\psi_{0}$ and $\psi_{4}$ more symmetric. In our notation, these rescalings are $\Phi_{2} = \psi_{0}$, \quad $\Phi_{1} = -\rho^{-1} \psi_{1}\sqrt{2}$,\quad $\Phi_{-1} = -\rho^{-3} \psi_{3}/\sqrt{2}$,\quad $\Phi_{-2} = \rho^{-4} \psi_{4}$. Since Chandrasekhar and Teukolsky [2] adopt the same null tetrad, the generating function $H$ is identical in both cases. Consequently, Eq. (3.10) yields the following rescalings of the Weyl scalars: $\Phi_{2} = \psi_{0}$,\quad $\Phi_{1} = -\rho^{-1} \psi_{1}$,\quad $\Phi_{-1} = -\rho^{-3} \psi_{3}$,\quad $\Phi_{-2} = \rho^{-4} \psi_{4}$, noting that $\chi_{2}^{(2)} \equiv \psi_{0}$, $\chi_{1}^{(2)} \equiv \psi_{1}$, $\chi_{-1}^{(2)} \equiv \psi_{3}$, and $\chi_{-2}^{(2)} \equiv \psi_{4}$. Clearly, except for constant multiplicative factors in the rescalings of $\psi_{1}$ and $\psi_{3}$, the Weyl scalar rescalings given by Eq. (3.10) agree completely with those introduced by Chandrasekhar in Ref. [56].

Therefore, Eq. (3.10) not only generalizes Chandrasekhar's rescaling transformation of the Weyl scalars to the wave functions for all massless particles with spin $s \leq 2$ on a generic type-D black hole background, but also reveals that the definition given by Eq. (3.10) is intimately related to the spin-weighted structure of the fields. Furthermore, since the spin coefficients $\rho$, $\mu$, $\tau$, and $\pi$ can be generated from $H$, it is natural to interpret $H$ as the generating function. It is worth noting that the transformation in Eq. (2.8) possesses the property of leaving the generating function $H$ invariant.

Using the defintion (3.1), the transformation (3.10) and the commutation relation [43],
\begin{eqnarray}\label{3.11}
&&\Delta D-D\Delta=(\gamma+\bar{\gamma})D+(\varepsilon+\bar{\varepsilon})\Delta-(\tau+\bar{\pi})\bar{\delta}-(\bar{\tau}+\pi)\delta, \nonumber\\
&&\bar{\delta}\delta-\delta\bar{\delta}=(\bar{\mu}-\mu)D+(\bar{\rho}-\rho)\Delta-(\bar{\alpha}-\beta)\bar{\delta}-(\bar{\beta}-\alpha)\delta,
\end{eqnarray}
along with the Newman-Penrose equations [43],
\begin{eqnarray}\label{3.12}
&&\Delta\rho-\bar{\delta}\tau=-(\rho\bar{\mu}+\sigma\lambda)+(\bar{\beta}-\alpha-\bar{\tau})\tau+(\gamma+\bar{\gamma})\rho+\nu\kappa-\psi_{2}-R/12, \nonumber\\
&&D\mu-\delta\pi=\bar{\rho}\mu+\sigma\lambda+\pi\bar{\pi}-(\varepsilon+\bar{\varepsilon})\mu-(\bar{\alpha}-\beta)\pi-\nu\kappa+\psi_{2}+R/12, \nonumber\\
&&D\gamma-\Delta\varepsilon=(\tau+\bar{\pi})\alpha+(\bar{\tau}+\pi)\beta-(\varepsilon+\bar{\varepsilon})\gamma-(\gamma+\bar{\gamma})\varepsilon+\tau\pi-\nu\kappa+\psi_{2}-R/24+\phi_{11}, \nonumber\\
&&\delta\alpha-\bar{\delta}\beta=\mu\rho-\lambda\sigma+\alpha\bar{\alpha}+\beta\bar{\beta}-2\alpha\beta+(\rho-\bar{\rho})\gamma+(\mu-\bar{\mu})\varepsilon-\psi_{2}+R/24+\phi_{11},
\end{eqnarray}
after a lengthy derivation, Eqs. (3.8) and (3.9) can be condensed into a single statement:
\begin{eqnarray}\label{3.13}
&&\big\{[D-(2p-1)\varepsilon+\bar{\varepsilon}-2p\rho-\bar{\rho}](\Delta-2p\gamma+\mu)
-[\delta+\bar{\pi}-\bar{\alpha}-(2p-1)\beta-2p\tau](\bar{\delta}+\pi-2p\alpha)\nonumber\\
&&-(2p-1)(p-1)\psi_2\big\}\Phi_{p}=\kappa_{s}T_{p},
\end{eqnarray}
where
\begin{eqnarray}\label{3.14}
T_{p}=H^{s-p} T_{p}^{(s)}.
\end{eqnarray}
Note that $\phi_{11}$ in Eq. (3.12) is one of the scalar functions that encode the tracefree Ricci tensor.
We know from Eq. (3.10) that the $-2s$th power of the generating function is a component of the wave function for the spin weight $p=-s$. When this component is ``peeled off'', the remaining part of the wave function satisfies the same dynamical equation as the wave function for the spin weight $p=s$.

For Kerr spacetime, substituting Teukolsky's null tetrad, spin coefficients, and the Weyl scalar $\psi_{2}$--as given in Ref. [3] (see also Eq. (A.6) in Appendix A)--into Eq. (3.13), the equation can be written as:
\begin{eqnarray}\label{3.15}
&&\{[\frac{(r^{2}+a^{2})^{2}}{{\sf\Delta}}-a^{2}\sin^{2}\theta]\frac{\partial^{2}}{\partial t^{2}}+\frac{4M a r}{{\sf\Delta}}\frac{\partial^{2}}{\partial t \partial \varphi}
+(\frac{a^{2}}{{\sf\Delta}}-\frac{1}{\sin^{2}\theta})\frac{\partial^{2}}{\partial\varphi^{2}} \nonumber\\
&&-{\sf\Delta}^{-p}\frac{\partial}{\partial r}({\sf\Delta}^{p+1}\frac{\partial}{\partial r})-\frac{1}{\sin\theta}\frac{\partial}{\partial\theta}(\sin\theta\frac{\partial}{\partial\theta})
-2p [\frac{a(r-M)}{{\sf\Delta}}+\frac{i\cos\theta}{\sin^{2}\theta}]\frac{\partial}{\partial\varphi} \nonumber\\
&&-2p[\frac{M(r^{2}-a^{2})}{{\sf\Delta}}-r-i a\cos\theta]\frac{\partial}{\partial t}
-p+p^{2}\cot^{2}\theta\}\Phi_{p} \nonumber\\
&&=2\kappa_{s}T_{p}(r^{2}+a^{2}\cos^{2}\theta).
\end{eqnarray}
Equation (3.15) is identical to Eq. (4.7) in Ref. [3], which provides an important consistency check.

To simplify Eq. (3.13), we introduce a quantity, $L_{\mu}$, which we call the spin-coefficient connection, and it is defined as:
\begin{eqnarray}\label{3.16}
L_{\mu}=2\lambda_{\mu i}Z^{i},
\end{eqnarray}
where
\begin{eqnarray}\label{3.17}
Z^{i}=(-\gamma,\,\, -\varepsilon-\rho,\,\, \alpha,\,\, \beta+\tau)^{T}.
\end{eqnarray}
Here $Z^{i}$ is the vector constructed using the spin coefficients. The spin-coefficient connection (3.16) can be expressed in a more intuitive form as follows:
\begin{equation}\label{3.18}
\begin{array}{ll}
\left(\begin{array}{lll}
L^{0}\\
L^{1}\\
L^{2}\\
L^{3}
\end{array}
\right)=2\left(\begin{array}{lll}
l^{0}\,\,\,\,\,\,n^{0}\,\,\,\,\,\,m^{0}\,\,\,\,\,\,\bar{m}^{0}\\
l^{1}\,\,\,\,\,\,n^{1}\,\,\,\,\,\,m^{1}\,\,\,\,\,\,\bar{m}^{1}\\
l^{2}\,\,\,\,\,\,n^{2}\,\,\,\,\,\,m^{2}\,\,\,\,\,\,\bar{m}^{2}\\
l^{3}\,\,\,\,\,\,n^{3}\,\,\,\,\,\,m^{3}\,\,\,\,\,\,\bar{m}^{3}
\end{array}
\right)
\left(\begin{array}{lll}
\,\,\,\,\,\,\,-\gamma\\
-\varepsilon-\rho\\
\,\,\,\,\,\,\,\,\,\,\alpha\\
\,\,\,\,\,\beta+\tau
\end{array}
\right).
\end{array}
\end{equation}

Using the orthonormal conditions of the null vectors, namely $\lambda^{\mu}\,_{i}\lambda_{\mu j}=\eta_{ij}$, the inner product $L^{\mu}L_{\mu}$ is easy to calculate:
\begin{eqnarray}\label{3.19}
L^{\mu}L_{\mu}=8[\gamma(\varepsilon+\rho)-\alpha(\beta+\tau)].
\end{eqnarray}

Applying the spin-coefficient connection (3.16), and after rather complicated calculations, Eq. (3.13) can be expressed in terms of the Weyl scalar $\psi_{2}$, and the Ricci scalar $R$:
\begin{equation}\label{3.20}
[(\nabla^{\mu}+pL^{\mu})(\nabla_{\mu}+pL_{\mu})-4p^{2}\psi_{2}+\frac{1}{6}R]\Phi_{p}=2\kappa_{s}T_{p}.
\end{equation}
Note that, as in Eq. (2.9), $\nabla_{\mu}$ denotes the covariant derivative in the metric $g_{\mu\nu}$. Evidently, when $p=0$ and $T_{p}=0$, Eq. (3.20) is just the (conformally invariant) massless scalar field equation. Therefore, Eq. (3.20) governs not only the massless fields of spin $1/2$, $1$, $3/2$, and $2$, but also the scalar field ($s=0$). We name Eq. (3.20) as the unified equation.

One of the key advantages of the Teukolsky master equation in Kerr spacetime is the separability of the angular and radial equations for $\Phi_{s}$ and $\Phi_{-s}$. Although a general discussion of the separability of the unified equation (3.20) for arbitrary Petrov type-D spacetimes remains challenging, we have verified that this important property persists for the Kerr-Newman-de Sitter spacetimes [57]. As a corollary, in special cases of these spacetimes--such as Kerr-de Sitter spacetime and Kerr spacetime--the unified equation naturally inherits the separability of the angular and radial equations for $\Phi_{s}$ and $\Phi_{-s}$. On the other hand,  Eq. (3.20) depends neither on specific coordinates nor on a particular type-D spacetime. Therefore, researchers can choose coordinates according to their needs in any type-D black hole spacetime and then apply Eq. (3.20) to study gravitational waves from EMRIs or to explore analogies between massless particle waves. Hence, Eq. (3.20) constitutes the fundamental equation for perturbation theory in type-D black hole spacetimes.

One often considers the source-free case, in which Eq. (3.20) is taken to be of the form
\begin{equation}\label{3.21}
[(\nabla^{\mu}+pL^{\mu})(\nabla_{\mu}+pL_{\mu})-4p^{2}\psi_{2}+\frac{1}{6}R]\Phi_{p}=0.
\end{equation}
It is surprising that the massless free-field equations for the nonzero spins $s\leq2$ have such a similar structure in any type-D black hole spacetime. Various field equations merge into a single unified equation, suggesting that these fields possess some common characteristics. We have verified this argument in two distinct scenarios: the Schwarzschild-type medium [58] and Grumiller spacetime [59].

\section{Verification of the new spin coefficient definitions and the explicit unified equation}
\label{sec:intro}

In Section 3.1, we demonstrate the validity of the newly defined spin coefficients in type-D vacuum spacetimes (with or without a cosmological constant). In this section, we prove the validity of Eq. (3.1) and present the explicit forms of the unified equation for several important black hole families. These include general spherically symmetric spacetimes, general Vaidya-type spacetimes, the Pleba$\acute{n}$ski-Demia$\acute{n}$ski metric, the complete family of black hole-like spacetimes, and the Kerr-Newman-de Sitter spacetime. All of these spacetimes are of Petrov type D and may contain electromagnetic fields, pure radiation, dust, perfect fluids, or other forms of matter.

\subsection{General spherically symmetric spacetimes}

A general metric for spherically symmetric spacetimes is given by
\begin{equation}\label{4.1}
ds^{2}=B(t, r)dt^{2}-A(t, r)dr^{2}-C(t, r)(d\theta^{2}+\sin^{2}\theta
d\varphi^{2}) \ .
\end{equation}
The metric (4.1) comprehensively describes all static and dynamic spherically symmetric black holes across a variety of theoretical frameworks, including Einstein's general relativity, loop quantum gravity, string theory, and modified gravitational theories. However, any given metric can, in principle, be represented as a solution to Einstein's equations by choosing a corresponding energy-momentum tensor. A partial catalog of black hole solutions included in Eq. (4.1) is provided in Table 1 of Appendix B.

We introduce a null tetrad of basis vectors, $l_{\mu}, n_{\mu}, m_{\mu}$, and $\bar{m}_{\mu}$, as follows:
\begin{eqnarray}\label{4.2}
&&l^{\mu}=(A, \sqrt{AB}, 0, 0),\nonumber\\
&&n^{\mu}=(\frac{1}{2AB}, -\frac{1}{2A\sqrt{AB}}, 0, 0),\nonumber\\
&&m^{\mu}=(0, 0, \frac{1}{\sqrt{2C}}, \frac{i}{\sqrt{2C}\sin\theta}),\nonumber\\
&&\bar{m}^{\mu}=(0, 0, \frac{1}{\sqrt{2C}}, -\frac{i}{\sqrt{2C}\sin\theta}).
\end{eqnarray}

Applying Eq. (4.2) and the Newman-Penrose definition (2.9), all spin coefficients can be computed (see Eq. (A.1) in Appendix A), where $\rho$, $\mu$, $\tau$, and $\pi$ are given by
\begin{eqnarray}\label{4.3}
&&\rho=-\frac{A}{2}\frac{\dot{C}}{C}-\frac{\sqrt{AB}}{2}\frac{C'}{C},
\quad \mu=\frac{1}{4\sqrt{AB}}(\frac{1}{\sqrt{AB}}\frac{\dot{C}}{C}-\frac{1}{A}\frac{C'}{C}), \quad \tau=\pi=0,
\end{eqnarray}
Here a prime denotes the derivative with respect to $r$, and a dot denotes the derivative with respect to $t$. Substituting Eqs. (4.2) and (4.3) into Eq. (3.3) and integrating yields
\begin{equation}\label{4.4}
H=\sqrt{C}.
\end{equation}
Note that the generating function $H$ derived from Eq. (3.3) naturally satisfies the new definition of spin coefficients in Eq. (3.1); thus, the existence of solution (4.4) confirms that Eq. (3.1) is valid for general spherically symmetric spacetimes.

The transformation of the wave function, derived from the generating function (4.4), takes the following form:
\begin{eqnarray}\label{4.5}
\chi_{p}^{(s)}=C^{(p-s)/2}\Phi_{p}.
\end{eqnarray}

The spin-coefficient connection $L^{\mu}$, Weyl scalar $\psi_{2}$, and Ricci scalar $R$ in Eq. (3.20) can be derived from Eqs. (3.2), (3.12), (3.18), (4.2), and (A.1), and are expressed as follows:
\begin{eqnarray}\label{4.6}
&&L^{0}=L^{t}=-\frac{1}{B}(\frac{\dot{A}}{A}+\frac{1}{2}\frac{\dot{B}}{B}-\frac{1}{2}\frac{\dot{C}}{C})-\frac{1}{2\sqrt{AB}}(\frac{B'}{B}-\frac{C'}{C}),\nonumber\\
&&L^{1}=L^{r}=\frac{1}{2\sqrt{AB}}(\frac{\dot{A}}{A}-\frac{\dot{C}}{C})+\frac{1}{A}(\frac{A'}{A}+\frac{1}{2}\frac{B'}{B}-\frac{1}{2}\frac{C'}{C}),\nonumber\\
&&L^{2}=L^{\theta}=0,\nonumber\\
&&L^{3}=L^{\varphi}=-\frac{i}{C}\frac{\cos\theta}{\sin^{2}\theta};
\end{eqnarray}
\begin{eqnarray}\label{4.7}
\psi_{2}&&=\frac{1}{24B}[\frac{\dot{A}\dot{B}}{AB}+\frac{\dot{A}\dot{C}}{AC}-\frac{\dot{B}\dot{C}}{BC}+(\frac{\dot{A}}{A})^{2}-2(\frac{\dot{C}}{C})^{2}-2\frac{\ddot{A}}{A}+2\frac{\ddot{C}}{C}]\nonumber\\
&&-\frac{1}{24A}[\frac{A'B'}{AB}-\frac{A'C'}{AC}+\frac{B'C'}{BC}+(\frac{B'}{B})^{2}-2(\frac{C'}{C})^{2}-2\frac{B''}{B}+2\frac{C''}{C}]-\frac{1}{6C};
\end{eqnarray}
and
\begin{eqnarray}\label{4.8}
R&&=\frac{1}{B}[-\frac{1}{2}\frac{\dot{A}\dot{B}}{AB}+\frac{\dot{A}\dot{C}}{AC}-\frac{\dot{B}\dot{C}}{BC}-\frac{1}{2}(\frac{\dot{A}}{A})^{2}-\frac{1}{2}(\frac{\dot{C}}{C})^{2}+\frac{\ddot{A}}{A}+2\frac{\ddot{C}}{C}]\nonumber\\
&&+\frac{1}{A}[\frac{1}{2}\frac{A'B'}{AB}+\frac{A'C'}{AC}-\frac{B'C'}{BC}+\frac{1}{2}(\frac{B'}{B})^{2}+\frac{1}{2}(\frac{C'}{C})^{2}-\frac{B''}{B}-2\frac{C''}{C}]+\frac{2}{C}.
\end{eqnarray}
Equations (4.6)-(4.8) provide the expressions for the quantities in the unified equation (3.20) based on the null tetrad (4.2). Consequently, the specific form of the unified equation in any spherically symmetric black hole-like spacetime can be obtained from these equations. For static spherically symmetric metric where $\dot{A}=\dot{B}=\dot{C}=0$, the spin coefficient (4.3), Weyl scalar (4.7), Ricci scalar (4.8), and spin-coefficient connection (4.6) reduce to the case discussed in Ref. [60].

\subsection{General Vaidya-type spacetimes}

In practice, it is often useful to introduce an advanced time coordinate $v$, which can help eliminate the problematic Schwarzschild time coordinate $t$. Using null coordinates, spherically symmetric metrics can thus be expressed in the form
\begin{equation}\label{4.9}
ds^{2}=A(v, r)dv^{2}-2B(v, r)dv dr-r^{2}(d\theta^{2}+\sin^{2}\theta
d\varphi^{2}),
\end{equation}
commonly known as a general metric for Vaidya-type spacetimes. It is worth noting that, for static spacetimes, there exists a relationship between the advanced time coordinate and the Schwarzschild time coordinate, expressed as $v = t + tortoise \,\, coordinate$. However, this relationship ceases to be valid when the metric function becomes time-dependent. The metric (4.9) can describe various spacetimes that utilize null coordinates; some of these are outlined in Table 2 of Appendix B.

Adopting the null tetrad [61]
\begin{eqnarray}\label{4.10}
&&l^{\mu}=(0, \frac{1}{B}, 0, 0),\nonumber\\
&&n^{\mu}=(-1, -\frac{A}{2B}, 0, 0),\nonumber\\
&&m^{\mu}=\frac{1}{\sqrt{2}r}(0, 0, 1, \frac{i}{\sin\theta}),\nonumber\\
&&\bar{m}^{\mu}=\frac{1}{\sqrt{2}r}(0, 0, 1, -\frac{i}{\sin\theta}).
\end{eqnarray}

The spin coefficients can be obtained from Eqs. (4.10) and (2.9), as given in Eq. (A.2) in Appendix A (see also Ref. [61]). From these results, we have
\begin{eqnarray}\label{4.11}
&&\rho=-\frac{1}{Br}, \quad \mu=-\frac{A}{2Br}, \quad \tau=\pi=0.
\end{eqnarray}
Inserting (4.10) and (4.11) into (3.3) and integrating yields
\begin{equation}\label{4.12}
H=r.
\end{equation}
The existence of a solution to Eq. (3.3) for general Vaidya-type spacetimes indicates that Eq. (3.1) is correct in such spacetimes.

The transformation of the wave function, as derived from the generating function (4.12), is expressed in the following form:
\begin{eqnarray}\label{4.13}
\chi_{p}^{(s)}=r^{p-s}\Phi_{p}.
\end{eqnarray}

The spin-coefficient connection $L^{\mu}$, Weyl scalar $\psi_{2}$, and Ricci scalar $R$ presented in Eq. (3.20) can be obtained from Eqs. (3.2), (3.12), (3.18), (4.10), and (A.2), and are formulated as follows:
\begin{eqnarray}\label{4.14}
&&L^{0}=L^{v}=-\frac{2}{Br},\nonumber\\
&&L^{1}=L^{r}=-\frac{1}{B^{2}}(\dot{B}+\frac{A'}{2}+\frac{A}{r}),\nonumber\\
&&L^{2}=L^{\theta}=0,\nonumber\\
&&L^{3}=L^{\varphi}=-\frac{i}{r^{2}}\frac{\cos\theta}{\sin^{2}\theta};
\end{eqnarray}
\begin{eqnarray}\label{4.15}
\psi_{2}=-\frac{B'}{6B^{3}}(\dot{B}+\frac{A'}{2}-\frac{A}{r})+\frac{1}{6B^{2}}[\dot{B}'+\frac{A''}{2}-\frac{A'}{r}-\frac{1}{r^{2}}(B^{2}-A)];
\end{eqnarray}
and
\begin{eqnarray}\label{4.16}
R=-\frac{B'}{B^{3}}(2\dot{B}+A'+4\frac{A}{r})+\frac{1}{B^{2}}(2\dot{B}'+A''+4\frac{A'}{r})-\frac{2}{r^{2}}(1-\frac{A}{B^{2}}).
\end{eqnarray}
Here the prime denotes the derivative with respect to $r$, and the dot denotes the derivative with respect to $v$. Equations (4.14)-(4.16) provide the expressions for the quantities in the unified equation (3.20). As a result, the explicit form of the unified equation in any Vaidya-type black hole spacetime can be derived from these equations.

\subsection{The Pleba$\acute{n}$ski-Demia$\acute{n}$ski metric}

The Pleba$\acute{n}$ski-Demia$\acute{n}$ski metric, as well as those derived from it via specific coordinate transformations, includes the complete family of Petrov type-D spacetimes with an aligned electromagnetic field and a potentially non-zero cosmological constant. In 2006, a modified version of the Pleba$\acute{n}$ski-Demia$\acute{n}$ski metric was introduced, allowing the most important special cases to be obtained via explicit reduction. The modified form of the metric is expressed as [62]
\begin{eqnarray}\label{4.17}
ds^{2}&&=\frac{1}{(1-\alpha \textsf{p}r)^{2}}[\frac{Q}{r^{2}+\omega^{2}\textsf{p}^{2}}(d\tau-\omega \textsf{p}^{2}d\sigma)^{2}-\frac{P}{r^{2}+\omega^{2}\textsf{p}^{2}}(\omega d\tau+r^{2}d\sigma)^{2} \nonumber\\
&&-\frac{r^{2}+\omega^{2}\textsf{p}^{2}}{P}d\textsf{p}^{2}-\frac{r^{2}+\omega^{2}\textsf{p}^{2}}{Q}dr^{2}],
\end{eqnarray}
where
\begin{equation}\label{4.18}
P=P(\textsf{p})=k+2\omega^{-1}n\textsf{p}-\epsilon \textsf{p}^{2}+2\alpha m\textsf{p}^{3}-[\alpha^{2}(\omega^{2}k+e^{2}+g^{2})+\omega^{2}\Lambda/3]\textsf{p}^{4},
\end{equation}
\begin{equation}\label{4.19}
Q=Q(r)=(\omega^{2}k+e^{2}+g^{2})-2mr+\epsilon r^{2}-2\alpha\omega^{-1}nr^{3}-(\alpha^{2}k+\Lambda/3)r^{4}.
\end{equation}
The parameters $m$, $n$, $e$, $g$, $\Lambda$ $\epsilon$, $k$, $\alpha$ and $\omega$ are arbitrary real values. It is important to note that, except for $\Lambda$, $e$ and $g$, the parameters in this metric do not necessarily retain their traditional physical interpretations; they only acquire well-defined meanings in specific sub-cases. Note that $\alpha$ and $\sigma$ in subsections 4.3 and 4.4 are not spin coefficients and differ from the convention introduced in section 1.

Based on the null tetrad established in [62], we have
\begin{eqnarray}\label{4.20}
&&l^{\mu}=\frac{1-\alpha \textsf{p}r}{\sqrt{2(r^{2}+\omega^{2}\textsf{p}^{2})Q}}(r^{2}, -Q, 0, -\omega),\nonumber\\
&&n^{\mu}=\frac{1-\alpha \textsf{p}r}{\sqrt{2(r^{2}+\omega^{2}\textsf{p}^{2})Q}}(r^{2}, Q, 0, -\omega),\nonumber\\
&&m^{\mu}=\frac{1-\alpha \textsf{p}r}{\sqrt{2(r^{2}+\omega^{2}\textsf{p}^{2})P}}(-\omega \textsf{p}^{2}, 0, iP, -1),\nonumber\\
&&\bar{m}^{\mu}=\frac{1-\alpha \textsf{p}r}{\sqrt{2(r^{2}+\omega^{2}\textsf{p}^{2})P}}(-\omega \textsf{p}^{2}, 0, -iP, -1).
\end{eqnarray}

Using Eqs. (4.20) and (2.9), we can compute the standard Newman-Penrose spin coefficients, and the result is Eq. (A.3) in Appendix A (see also Ref. [62]). From these, we have
\begin{eqnarray}\label{4.21}
&&\rho=\mu=\sqrt{\frac{Q}{2(r^{2}+\omega^{2}\textsf{p}^{2})}}\frac{1+i\alpha\omega \textsf{p}^{2}}{r+i\omega \textsf{p}},\nonumber\\
&&\tau=\pi=\sqrt{\frac{P}{2(r^{2}+\omega^{2}\textsf{p}^{2})}}\frac{\omega-i\alpha r^{2}}{r+i\omega \textsf{p}}.
\end{eqnarray}
When these relations together with (4.20) are substituted into (3.3) and the result is integrated, we obtain
\begin{equation}\label{4.22}
H=\frac{r+i \omega \textsf{p}}{1-\alpha \textsf{p} r}.
\end{equation}
The existence of the generating function Eq. (4.22) confirms that the new definition of the spin coefficients, Eq. (3.1), is valid for the Pleba$\acute{n}$ski-Demia$\acute{n}$ski metric.

The transformation of the wave function, derived from the generating function (4.22), is expressed as follows:
\begin{eqnarray}\label{4.23}
\chi_{p}^{(s)}=(\frac{r+i \omega \textsf{p}}{1-\alpha \textsf{p} r})^{p-s}\Phi_{p}.
\end{eqnarray}

The spin-coefficient connection $L^{\mu}$, Weyl scalar $\psi_{2}$, and Ricci scalar $R$ in Eq. (3.20) can be derived from Eqs. (3.2), (3.12), (3.18), (4.20), and (A.3), and are expressed in terms of
\begin{eqnarray}\label{4.24}
&&L^{0}=L^{\tau}=-\frac{2(1-\alpha \textsf{p}r)^{2}}{r+i\omega \textsf{p}}+\frac{(1-\alpha \textsf{p}r)^{2}}{2(r^{2}+\omega^{2} \textsf{p}^{2})}(r^{2}\frac{\partial_{r}Q}{Q}-i\omega \textsf{\textsf{p}}^{2}\frac{\partial_{\textsf{p}}P}{P}),\nonumber\\
&&L^{1}=L^{r}=-\frac{Q(1-\alpha \textsf{p}r)(1+i\alpha\omega \textsf{p}^{2})}{(r+i\omega \textsf{p})(r^{2}+\omega^{2} \textsf{p}^{2})},\nonumber\\
&&L^{2}=L^{\textsf{p}}=-i\frac{P(1-\alpha \textsf{p}r)(\omega-i\alpha r^{2})}{(r+i\omega \textsf{p})(r^{2}+\omega^{2} \textsf{p}^{2})},\nonumber\\
&&L^{3}=L^{\sigma}=-\frac{(1-\alpha \textsf{p}r)^{2}}{2(r^{2}+\omega^{2} \textsf{p}^{2})}(\omega\frac{\partial_{r}Q}{Q}+i\frac{\partial_{\textsf{p}}P}{P});
\end{eqnarray}
\begin{eqnarray}\label{4.25}
\psi_{2}=-(m+in)(\frac{1-\alpha \textsf{p}r}{r+i\omega \textsf{p}})^{3}+(e^{2}+g^{2})(\frac{1-\alpha \textsf{p}r}{r+i\omega \textsf{p}})^{3}\frac{1+\alpha \textsf{p}r}{r-i\omega \textsf{p}};
\end{eqnarray}
and
\begin{eqnarray}\label{4.26}
R=4\Lambda.
\end{eqnarray}
Equations (4.24)-(4.26) present the explicit form of the unified equation in Pleba$\acute{n}$ski-Demia$\acute{n}$ski spacetimes.

\subsection{The complete family of black hole-like spacetimes}

In certain limits, by making coordinate transformations, the Pleba$\acute{n}$ski-Demia$\acute{n}$ski metric can become the following complete family of black hole-like solutions [62]:
\begin{equation}\label{4.27}
ds^{2}=\frac{1}{\Omega^{2}}\{\frac{Q}{\varrho^{2}}[dt-(a\sin^{2}\theta+4l\sin^{2}\frac{\theta}{2})d\varphi]^{2}-\frac{\varrho^{2}}{Q}dr^{2}-\frac{\widetilde{P}}{\varrho^{2}}[adt-(r^{2}+(a+l)^{2})d\varphi]^{2}-\frac{\varrho^{2}}{\widetilde{P}}\sin^{2}\theta d\theta^{2}\},
\end{equation}
where
\begin{eqnarray}\label{4.28}
&&\Omega=1-\frac{\alpha}{\omega}(l+a\cos\theta)r,\nonumber\\
&&\varrho=r^{2}+(l+a\cos\theta)^{2},\nonumber\\
&&\widetilde{P}=\sin^{2}\theta(1-a_{3}\cos\theta-a_{4}\cos^{2}\theta),\nonumber\\
&&Q=(\omega^{2}k+e^{2}+g^{2})-2mr+\epsilon r^{2}-2\alpha\frac{n}{\omega}r^{3}-(\alpha^{2}k+\frac{\Lambda}{3})r^{4},
\end{eqnarray}
and
\begin{eqnarray}\label{4.29}
&&a_{3}=2\alpha\frac{a}{\omega}m-4\alpha^{2}\frac{al}{\omega^{2}}(\omega^{2}k+e^{2}+g^{2})-4\frac{\Lambda}{3}al,\nonumber\\
&&a_{4}=-\alpha^{2}\frac{a^{2}}{\omega^{2}}(\omega^{2}k+e^{2}+g^{2})-\frac{\Lambda}{3}a^{2},
\end{eqnarray}
with
\begin{eqnarray}\label{4.30}
&&\epsilon=\frac{\omega^{2}k}{a^{2}-l^{2}}+4\alpha\frac{l}{\omega}m-(a^{2}+3l^{2})[\frac{\alpha^{2}}{\omega^{2}}(\omega^{2}k+e^{2}+g^{2})+\frac{\Lambda}{3}],\nonumber\\
&&n=\frac{\omega^{2}kl}{a^{2}-l^{2}}-\alpha\frac{a^{2}-l^{2}}{\omega}m+(a^{2}-l^{2})l[\frac{\alpha^{2}}{\omega^{2}}(\omega^{2}k+e^{2}+g^{2})+\frac{\Lambda}{3}],\nonumber\\
&&k=[1+2\alpha\frac{l}{\omega}m-3\alpha^{2}\frac{l^{2}}{\omega^{2}}(e^{2}+g^{2})-l^{2}\Lambda](\frac{\omega^{2}}{a^{2}-l^{2}}+3\alpha^{2}l^{2})^{-1}.
\end{eqnarray}
Metric (4.27) contains eight arbitrary constants: the mass parameter $m$ of the source, its electric charge $e$, magnetic charge $g$, Kerr-like rotation parameter $a$, NUT parameter $l$, acceleration $\alpha$, and cosmological constant $\Lambda$. Additionally, there is the parameter $\omega$, which can be set to any convenient value if either $a$ or $l$ is nonzero; otherwise, $\omega\equiv 0$. The complete family of black hole-like metrics covers many well-known black hole spacetimes, with the partial black hole solutions listed in Table 3 of Appendix B.

The null tetrad can be chosen as
\begin{eqnarray}\label{4.31}
&&l^{\mu}=\frac{\Omega}{\sqrt{2Q}\varrho}\{[r^{2}+(l+a)^{2}], -Q, 0, a\},\nonumber\\
&&n^{\mu}=\frac{\Omega}{\sqrt{2Q}\varrho}\{[r^{2}+(l+a)^{2}], Q, 0, a\},\nonumber\\
&&m^{\mu}=\frac{\Omega}{\sqrt{2\widetilde{P}}\varrho}\{\frac{1}{a}[(l+a)^{2}-(l+a\cos\theta)^{2}], 0, -i\frac{\widetilde{P}}{\sin\theta}, 1\},\nonumber\\
&&\bar{m}^{\mu}=\frac{\Omega}{\sqrt{2\widetilde{P}}\varrho}\{\frac{1}{a}[(l+a)^{2}-(l+a\cos\theta)^{2}], 0, i\frac{\widetilde{P}}{\sin\theta}, 1\}.
\end{eqnarray}

From Eqs. (4.31) and (2.9) we can compute the spin coefficients, with the complete expressions collected in Eq. (A.4) of Appendix A. In particular, the spin coefficients are given by
\begin{eqnarray}\label{4.32}
&&\rho=\mu=\sqrt{\frac{Q}{2}}\frac{1+i\frac{\alpha}{\omega}(l+a\cos\theta)^{2}}{\varrho[r+i(l+a\cos\theta)]},\nonumber\\
&&\tau=\pi=\sqrt{\frac{\widetilde{P}}{2}}\frac{a(1-i\frac{\alpha}{\omega}r^{2})}{\varrho[r+i(l+a\cos\theta)]}.
\end{eqnarray}
Substituting (4.31) and (4.32) into (3.3) and carrying out the integration leads to
\begin{equation}\label{4.33}
H=\frac{r+i(l+a\cos\theta)}{1-\frac{\alpha}{\omega}(l+a\cos\theta)r}.
\end{equation}
The existence of a solution to Eq. (3.3) for the complete family of black hole-like spacetimes indicates that Eq. (3.1) holds there as well.

The transformation of the wave function, determined by the generating function (4.33), can be expressed as
\begin{eqnarray}\label{4.34}
\chi_{p}^{(s)}=(\frac{r+i(l+a\cos\theta)}{1-\frac{\alpha}{\omega}(l+a\cos\theta)r})^{p-s}\Phi_{p}.
\end{eqnarray}

The spin-coefficient connection $L^{\mu}$, the Weyl scalar $\psi_{2}$, and the Ricci scalar $R$ presented in Eq. (3.20) can be derived from Eqs. (3.2), (3.12), (3.18), (4.31), and (A.4). These quantities can be expressed in the form
\begin{eqnarray}\label{4.35}
&&L^{0}=L^{t}=-\frac{2\Omega^{2}}{r+i(l+a\cos\theta)}+\frac{\Omega^{2}}{2\varrho^{2}}\{[r^{2}+(l+a)^{2}]\frac{\partial_{r}Q}{Q}-[(l+a)^{2}-(l+a\cos\theta)^{2}]\frac{i}{\sin\theta}\frac{\partial_{\theta}\widetilde{P}}{\widetilde{P}}\},\nonumber\\
&&L^{1}=L^{r}=-\frac{Q\Omega[1+i\frac{\alpha}{\omega}(l+a\cos\theta)^{2}]}{\varrho^{2}[r+i(l+a\cos\theta)]},\nonumber\\
&&L^{2}=L^{\theta}=\frac{ia\Omega \widetilde{P}(1-i\frac{\alpha}{\omega}r^{2})}{\varrho^{2}\sin\theta[r+i(l+a\cos\theta)]},\nonumber\\
&&L^{3}=L^{\varphi}=\frac{\Omega^{2}}{2\varrho^{2}}(a\frac{\partial_{r}Q}{Q}-\frac{i}{\sin\theta}\frac{\partial_{\theta}\widetilde{P}}{\widetilde{P}}\};
\end{eqnarray}
\begin{eqnarray}\label{4.36}
\psi_{2}=(\frac{1-\frac{\alpha}{\omega}(l+a\cos\theta)r}{r+i(l+a\cos\theta)})^{3}[-(m+in)+(e^{2}+g^{2})\frac{1+\frac{\alpha}{\omega}(l+a\cos\theta)r}{r-i(l+a\cos\theta)}];
\end{eqnarray}
and
\begin{eqnarray}\label{4.37}
R=4\Lambda.
\end{eqnarray}
The explicit form of the unified equation in complete family of black hole-like spacetimes is presented in Eqs. (4.35)-(4.37).

\subsection{The Kerr-Newman-de Sitter spacetime}

In the previous subsection, in the spacetime of a complete family of black hole-like metrics, by using the null tetrad (4.31) we found the spin coefficients: $\rho=\mu$ and $\tau=\pi$. In this subsection, in the context of the Kerr-Newman-de Sitter spacetime, we adopt a null tetrad of a type distinctly different from that in Eq. (4.31) to further validate the new definition of the spin coefficient presented in Eq. (3.1). Additionally, we provide an explicit formulation of the unified equation (3.20).

The Kerr-Newman-de Sitter spacetime can be expressed in Boyer-Lindquist-type coordinates as [44]
\begin{equation}\label{4.38}
ds^{2}=\frac{\rho\bar{\rho}\Delta_{r}}{\Xi^{2}}(dt-a\sin^{2}\theta d\varphi)^{2}-\frac{\rho\bar{\rho}\Delta_{\theta}\sin^{2}\theta}{\Xi^{2}}[adt-(r^{2}+a^{2})d\varphi]^{2}-\frac{1}{\rho\bar{\rho}}(\frac{dr^{2}}{\Delta_{r}}+\frac{d\theta^{2}}{\Delta_{\theta}}),
\end{equation}
where
\begin{eqnarray}\label{4.39}
&&\rho=-\frac{1}{r-ia\cos\theta},\nonumber\\
&&\Delta_{r}=(r^{2}+a^{2})(1-\frac{\Lambda}{3}r^{2})-2Mr+Q^{2},\nonumber\\
&&\Delta_{\theta}=1+\frac{\Lambda}{3}a^{2}\cos^{2}\theta,\nonumber\\
&&\Xi=1+\frac{\Lambda}{3}a^{2}.
\end{eqnarray}
Here $\Lambda$ is the cosmological constant, $M$ is the mass of the black hole, $Q$ is its charge and $a$ is the angular momentum per unit mass. Note that only when the parameters $m\neq 0$ $e\neq 0$, $a\neq 0$, and $\Lambda\neq 0$ does the complete family of black hole-like metrics (4.27) reduce to the Kerr-Newman-de Sitter metric, which can be expressed in the standard form of the Kerr-Newman-de Sitter solution (4.38) through the substitutions $t\rightarrow t \Xi^{-1}$ and $\varphi\rightarrow\varphi\Xi^{-1}$.

The null tetrad is chosen as [44]:
\begin{eqnarray}\label{4.40}
&&l^{\mu}=[\frac{(r^{2}+a^{2})\Xi}{\Delta_{r}}, 1, 0, \frac{a\Xi}{\Delta_{r}}],\nonumber\\
&&n^{\mu}=\frac{\rho\bar{\rho}}{2}[(r^{2}+a^{2})\Xi, -\Delta_{r}, 0, a\Xi],\nonumber\\
&&m^{\mu}=-\frac{\bar{\rho}}{\sqrt{2\Delta_{\theta}}}[ia\Xi\sin\theta, 0, \Delta_{\theta}, \frac{i\Xi}{\sin\theta}],\nonumber\\
&&\bar{m}^{\mu}=-\frac{\rho}{\sqrt{2\Delta_{\theta}}}[-ia\Xi\sin\theta, 0, \Delta_{\theta}, -\frac{i\Xi}{\sin\theta}].
\end{eqnarray}

Using Eqs. (4.40) and (2.9) we can obtain the spin coefficients; the full results are given in Eq. (A.5) of Appendix A (see also Ref. [44]). From these results, we have
\begin{eqnarray}\label{4.41}
\rho=-\frac{1}{r-ia\cos\theta}, \quad \mu=\frac{1}{2}\rho^{2}\bar{\rho}\Delta_{r}, \quad \tau=-i\sqrt{\frac{\Delta_{\theta}}{2}}\rho\bar{\rho} a\sin\theta, \quad
\pi=i\sqrt{\frac{\Delta_{\theta}}{2}}\rho^{2}a\sin\theta. \nonumber\\
\end{eqnarray}
Note that, unlike Eq. (4.32), here $\rho\neq\mu$ and $\tau\neq\pi$. After substituting (4.40) and (4.41) into (3.3) and integrating, we arrive at
\begin{equation}\label{4.42}
H=r-ia\cos\theta.
\end{equation}
From the existence of a solution to Eq. (3.3) for the Kerr-Newman-de Sitter spacetime, it follows that Eq. (3.1) is correct there.

The transformation of the wave function, as determined by the generating function in Eq. (4.42), can be expressed as:
\begin{eqnarray}\label{4.43}
\chi_{p}^{(s)}=(r-ia\cos\theta)^{p-s}\Phi_{p}.
\end{eqnarray}

The spin-coefficient connection $L^{\mu}$, the Weyl scalar $\psi_{2}$, and the Ricci scalar $R$ given in Eq. (3.20) can be derived from Eqs. (3.2), (3.12), (3.18), (4.40), and (A.5). These quantities can be expressed in the following form:
\begin{eqnarray}\label{4.44}
&&L^{0}=L^{t}=-\rho\bar{\rho} \Xi[(r^{2}+a^{2})\frac{\Delta_{r}'}{2 \Delta_{r}}+\frac{2}{\bar{\rho}}+ia\sin\theta(\frac{\Delta_{\theta}'}{2 \Delta_{\theta}}+\cot\theta)],\nonumber\\
&&L^{1}=L^{r}=-\frac{1}{2}\rho\bar{\rho} \Delta_{r}',\nonumber\\
&&L^{2}=L^{\theta}=0,\nonumber\\
&&L^{3}=L^{\varphi}=-\rho\bar{\rho}\Xi[a\frac{\Delta_{r}'}{2\Delta_{r}}+\frac{i}{\sin\theta}(\frac{\Delta_{\theta}'}{2\Delta_{\theta}}+\cot\theta)];
\end{eqnarray}
\begin{eqnarray}\label{4.45}
\psi_{2}=\rho^{3}(M+\bar{\rho}Q^{2});
\end{eqnarray}
and
\begin{eqnarray}\label{4.46}
R=4\Lambda,
\end{eqnarray}
where the prime denotes the derivative with respect to the independent variable. The explicit form of the unified equation in Kerr-Newman-de Sitter spacetimes is given by Eqs. (4.44)-(4.46).

Using Eqs. (4.44)-(4.46), Eq. (3.20) takes the following form in Kerr-Newman-de Sitter spacetime:
\begin{eqnarray}\label{4.47}
&&\{\Xi^{2}[\frac{(r^{2}+a^{2})^{2}}{\Delta_{r}}-\frac{a^{2}\sin^{2}\theta}{\Delta_{\theta}}]\frac{\partial^{2}}{\partial t^{2}}+2a \Xi^{2}(\frac{r^{2}+a^{2}}{\Delta_{r}}-\frac{1}{\Delta_{\theta}})\frac{\partial^{2}}{\partial t \partial \varphi} \nonumber\\
&&+\Xi^{2}(\frac{a^{2}}{\Delta_{r}}-\frac{1}{\Delta_{\theta} \sin^{2}\theta})\frac{\partial^{2}}{\partial\varphi^{2}}-\Delta_{r}^{-p}\frac{\partial}{\partial r}(\Delta_{r}^{p+1}\frac{\partial}{\partial r}) \nonumber\\
&&-\frac{1}{\sin\theta}\frac{\partial}{\partial\theta}(\Delta_{\theta}\sin\theta\frac{\partial}{\partial\theta})-2p \Xi[\frac{a\Delta_{r}'}{2\Delta_{r}}+\frac{i}{\sin\theta}(\frac{\Delta_{\theta}'}{2\Delta_{\theta}}+\cot\theta)]\frac{\partial}{\partial\varphi} \nonumber\\
&&-2p \Xi[i a\sin\theta(\frac{\Delta_{\theta}'}{2\Delta_{\theta}}-\cot\theta)+\frac{\Delta_{r}'}{2\Delta_{r}}(r^{2}+a^{2})-2r]\frac{\partial}{\partial t} \nonumber\\
&&-\frac{1}{2}p\Delta_{r}''+p^{2}\Delta_{\theta}(\frac{\Delta_{\theta}'}{2\Delta_{\theta}}+\cot\theta)^{2}+\frac{2}{3}\Lambda(2p^{2}+1)(r^{2}+a^{2}\cos^{2}\theta)\}\Phi_{p} \nonumber\\
&&=2\kappa_{s}T_{p}(r^{2}+a^{2}\cos^{2}\theta).
\end{eqnarray}
Under $T_{p}=0$, Eq. (4.47) is identical to Eq. (13) in Ref. [44]. When $ Q = 0 $ and $ \Lambda = 0 $, Eq. (4.47) reduces to the Teukolsky master equation [3]. This serves as a consistency check for the unified equation (3.20).

\section{Conclusion}
\label{sec:intro}

We have introduced a new definition (3.1) for the spin coefficients $\rho$, $\mu$, $\tau$, and $\pi$, which are interpreted as the rate of change of the logarithm of the generating function along the null tetrad ($l^{\mu}$, $n^{\mu}$, $m^{\mu}$, $\bar{m}^{\mu}$), respectively. It is important to note that the new definition of spin coefficients is applicable not only to Petrov type-D spacetimes but also to non-Petrov type-D spacetimes. For example, in a variable-mass Kerr spacetime, only Weyl tensor components $\psi_{0}$ and $\psi_{1}$ vanish [63, 64]. This indicates that the spacetime admits one principal null direction of multiplicity 2, while the others are distinct. Consequently, it is algebraically special of Petrov type II. The metric takes the form
\begin{eqnarray}\label{5.1}
ds^{2}&&=[1-2M(v)r\varrho\bar{\varrho}]dv^{2}-2dvdr+4M(v)ra\varrho\bar{\varrho}\sin^{2}\theta dvd\varphi+2a\sin^{2}\theta drd\varphi \nonumber\\
&&-(\varrho\bar{\varrho})^{-1}d\theta^{2}-[2M(v)ra^{2}\varrho\bar{\varrho}\sin^{2}\theta+r^{2}+a^{2}]\sin^{2}\theta d\varphi^{2},
\end{eqnarray}
where
\begin{eqnarray}\label{5.2}
\varrho=-\frac{1}{r-ia\cos\theta}.
\end{eqnarray}
We choose [61]
\begin{eqnarray}\label{5.3}
&&l^{\mu}=[0, 1, 0, 0],\nonumber\\
&&n^{\mu}=-\varrho\bar{\varrho}[(r^{2}+a^{2}), \frac{1}{2}(r^{2}+a^{2}-2Mr), 0, a],\nonumber\\
&&m^{\mu}=-\frac{\bar{\varrho}}{\sqrt{2}}[ia\sin\theta, 0, 1, \frac{i}{\sin\theta}],\nonumber\\
&&\bar{m}^{\mu}=-\frac{\varrho}{\sqrt{2}}[-ia\sin\theta, 0, 1, -\frac{i}{\sin\theta}],
\end{eqnarray}
and assume the generating function
\begin{eqnarray}\label{5.4}
H=r+i a \cos\theta.
\end{eqnarray}
Substituting these two equations into Eq. (3.1), we obtain
\begin{eqnarray}\label{5.5}
\rho=\bar{\varrho}, \quad \mu=\frac{\varrho\bar{\varrho}^{2}}{2}(r^{2}+a^{2}-2Mr), \quad \tau=\frac{\bar{\varrho}^{2}}{\sqrt{2}}ia\sin\theta, \quad \pi=-\frac{\varrho\bar{\varrho}}{\sqrt{2}}ia\sin\theta.
\end{eqnarray}
These spin coefficients are identical to those obtained from Eq. (2.9) (see Ref. [61] or Eq. (A.7) of Appendix A).

In summary, the spin coefficient definition (3.1) is firmly established and holds without exception in all type-D spacetimes. For non-type D spacetimes, however, it remains a working hypothesis, as no counterexamples have been found to date. On the other hand, the Bianchi identities tell us that, for type-D vacuum spacetimes, the generating function $H$ in Eq. (3.1) satisfies $H \propto (\psi_{2})^{-1/3}$. However, as seen from Eqs. (4.4) and (4.7), (4.12) and (4.15), (4.22) and (4.25), (4.33) and (4.36), as well as (4.42) and (4.45), $H$ is generally not a function of $\psi_{2}$ for type-D non-vacuum spacetimes. Therefore, compared with the formulations in existing literature [2, 3], and [56]--which directly parameterize rescaled quantities in terms of $\rho$ (for Kerr metric, $\psi_{2}\sim\rho^{3}$)--the use of $H$ offers both greater generality and additional technical advantages.

We have introduced a new concept: the spin coefficient connection $L_{\mu}$, as defined by Eq. (3.16). Employing this and the new spin coefficient definitions, we found that all massless fields with spin $s\leq2$ obey a single unified equation (3.20) in Petrov type-D spacetimes. As shown in Refs. [58, 59], the unified equation allows the wave functions of all particles to be determined simultaneously. This not only facilitates the investigation of individual particle properties but also allows for an exploration of the analogous characteristics shared between different types of particles. Since the properties of particles bear the imprint of their spacetime background, they provide insights into the nature of space and time and help us understand the features and behavior of phenomena where black holes are wholly or partially composed of curved spacetime.

In particular, modeling EMRIs primarily relies on black hole perturbation theory [4]. During the inspiral, the motion of the stellar compact object generates persistent gravitational radiation.
The unified equation (3.20) exhibits separability of the angular and radial equations for $\Phi_{s}$ and $\Phi_{-s}$ in at least several astrophysically important type-D black hole spacetimes, such as Kerr-Newman-de Sitter, Kerr-de Sitter, and Kerr spacetimes. This property, analogous to that of the Teukolsky master equation in Kerr spacetime, enables modeling EMRIs as perturbations on such a type-D black hole background generated by a supermassive black hole.

As mentioned in the introduction, in the weak-field approximation, the Einstein field equation can be reduced to equations similar to Maxwell's equations, thereby establishing a specialized theory known as gravitoelectromagnetism. The unified equation we have found here depends neither on the strength of the gravitational field nor on the specific metric form and coordinate system, and is universally applicable to any type-D black hole.

Note that the spin-coefficient connection contains the spin coefficient $\rho$. According to the geometrical interpretation of $\rho$, its real and imaginary parts correspond, respectively (minus), to the expansion and twist of the congruence of integral curves of $l^{\mu}$. Therefore, the first term within the square bracket in Eq. (3.20) indicates that contraction and rotation occur as the wave functions evolve from one point to another. On the other hand, Penrose [65] points out that the Weyl tensor acts as a purely astigmatic lens, while the Ricci scalar is proportional to a cosmological constant for usual black hole solutions [62]. A positive cosmological constant contributes repulsively to gravitational effects. Roughly speaking, the second and third terms within the brackets in Eq. (3.20) lead to wave functions that exhibit convergence and divergence during propagation processes.

We have validated the new definition of spin coefficients and applied the unified equation to general spherically symmetric spacetimes, general Vaidya-type spacetimes, the Pleba$\acute{n}$ski-Demia$\acute{n}$ski metric, the complete family of black hole-like spacetimes, and the Kerr-Newman-de Sitter spacetime, deriving specific expressions for each term in the equation. Since these metrics nearly encompass all known type-D black holes, this indicates that we have provided the explicit form of the unified equation governing massless spin particles in the background of every known type-D black hole. This lays a solid foundation for studying gravitational waves from EMRIs and for exploring the analogy between massless spin-particle waves in any type-D black hole background. For instance, using Eqs. (3.21) and (4.5)-(4.8), Ref. [58] investigated the analogy among such particles through scattering in a Schwarzschild-type medium, while Ref. [59] studied it via quasinormal modes in Grumiller spacetime.

\appendix
\section{Spin coefficients for six metrics}

All the spin coefficients in this appendix are calculated according to the Newman-Penrose definition (2.9) and the null tetrad given in sections 4 and 5.

\textit{1. A general metric for spherically symmetric spacetimes}
\begin{eqnarray}\label{A.1}
&&\kappa=\sigma=\nu=\lambda=\pi=\tau=0,\nonumber\\
&&\rho=-\frac{A}{2}\frac{\dot{C}}{C}-\frac{\sqrt{AB}}{2}\frac{C'}{C},
\quad \mu=\frac{1}{4\sqrt{AB}}(\frac{1}{\sqrt{AB}}\frac{\dot{C}}{C}-\frac{1}{A}\frac{C'}{C}), \quad \alpha=-\beta=-\frac{1}{2\sqrt{2C}}\cot\theta ,\nonumber\\
&&\varepsilon=\frac{A}{4}(3\frac{\dot{A}}{A}+\frac{\dot{B}}{B})+\frac{\sqrt{AB}}{2}(\frac{A'}{A}+\frac{B'}{B}), \quad \gamma=\frac{1}{8AB}(\frac{\dot{A}}{A}+\frac{\dot{B}}{B})-\frac{1}{4A\sqrt{AB}}\frac{A'}{A}.
\end{eqnarray}

\textit{2. A general metric for Vaidya-type spacetimes}
\begin{eqnarray}\label{A.2}
&&\kappa=\sigma=\nu=\lambda=\pi=\tau=\varepsilon=0,\nonumber\\
&&\rho=-\frac{1}{Br},
\quad \mu=-\frac{A}{2Br}, \quad \alpha=-\beta=-\frac{1}{2\sqrt{2}r}\cot\theta , \quad \gamma=\frac{\dot{B}}{2B}+\frac{A'}{4B}.
\end{eqnarray}

\textit{3. A modified form Of the Pleba$\acute{n}$ski-Demia$\acute{n}$ski metric}
\begin{eqnarray}\label{A.3}
&&\kappa=\sigma=\nu=\lambda=0,\nonumber\\
&&\rho=\mu=\sqrt{\frac{Q}{2(r^{2}+\omega^{2}\textsf{p}^{2})}}\frac{1+i\alpha\omega \textsf{p}^{2}}{r+i\omega \textsf{p}},\nonumber\\
&&\tau=\pi=\sqrt{\frac{P}{2(r^{2}+\omega^{2}\textsf{p}^{2})}}\frac{\omega-i\alpha r^{2}}{r+i\omega \textsf{p}},\nonumber\\
&&\varepsilon=\gamma=\frac{1}{4}\sqrt{\frac{Q}{2(r^{2}+\omega^{2}\textsf{p}^{2})}}[2\frac{1-\alpha \textsf{p}r}{r+i\omega \textsf{p}}-2\alpha \textsf{p}-(1-\alpha \textsf{p}r)\frac{\partial_{r}Q}{Q}],\nonumber\\
&&\alpha=\beta=\frac{1}{4}\sqrt{\frac{P}{2(r^{2}+\omega^{2}\textsf{p}^{2})}}[2\omega\frac{1-\alpha \textsf{p}r}{r+i\omega \textsf{p}}+2i\alpha r+i(1-\alpha \textsf{p}r)\frac{\partial_{\textsf{p}}P}{P}].
\end{eqnarray}
Note that in eq. (A.3), the $\alpha$ on the right-hand side is a real parameter.

\textit{4. The complete family of black hole-like spacetimes}
\begin{eqnarray}\label{A.4}
&&\kappa=\sigma=\nu=\lambda=0,\nonumber\\
&&\rho=\mu=\sqrt{\frac{Q}{2}}\frac{1+i\frac{\alpha}{\omega}(l+a\cos\theta)^{2}}{\varrho[r+i(l+a\cos\theta)]},\nonumber\\
&&\tau=\pi=\sqrt{\frac{\widetilde{P}}{2}}\frac{a(1-i\frac{\alpha}{\omega}r^{2})}{\varrho[r+i(l+a\cos\theta)]},\nonumber\\
&&\varepsilon=\gamma=\frac{1}{4\varrho}\sqrt{\frac{Q}{2}}[\frac{2\Omega}{r+i(l+a\cos\theta)}-\frac{2\alpha}{\omega}(l+a\cos\theta)-\Omega\frac{\partial_{r}Q}{Q}],\nonumber\\
&&\alpha=\beta=\frac{1}{4\varrho}\sqrt{\frac{\widetilde{P}}{2}}[\frac{2a\Omega}{r+i(l+a\cos\theta)}+i\frac{2\alpha}{\omega}ar-i\frac{\Omega}{\sin\theta}\frac{\partial_{\theta}\widetilde{P}}{\widetilde{P}}].
\end{eqnarray}
Note that in eq. (A.4), the $\alpha$ on the right-hand side is an acceleration parameter.

\textit{5. The Kerr-Newman-de Sitter spacetime}
\begin{eqnarray}\label{A.5}
&&\kappa=\sigma=\nu=\lambda=\varepsilon=0,\nonumber\\
&&\rho=-\frac{1}{r-ia\cos\theta}, \quad \tau=-i\sqrt{\frac{\Delta_{\theta}}{2}}\rho\bar{\rho} a\sin\theta, \quad \mu=\frac{1}{2}\rho^{2}\bar{\rho}\Delta_{r}, \quad \gamma=\frac{\rho\bar{\rho}}{4}\Delta_{r}'+\mu, \nonumber\\
&&\pi=i\sqrt{\frac{\Delta_{\theta}}{2}}\rho^{2}a\sin\theta, \quad \beta=-\frac{\sqrt{\Delta_{\theta}}\bar{\rho}}{2\sqrt{2}}(\frac{\Delta_{\theta}'}{2\Delta_{\theta}}+\cot\theta), \quad \alpha=\pi-\bar{\beta}.
\end{eqnarray}
When $Q=0$ and $\Lambda=0$, the Kerr-Newman-de Sitter metric reduces to the Kerr metric, and Eqs. (4.40), (A.5), (4.45), and (4.46) reduce accordingly to:
\begin{eqnarray}\label{A.6}
&&l^{\mu}=[\frac{(r^{2}+a^{2})}{{\sf\Delta}}, 1, 0, \frac{a}{{\sf\Delta}}],\nonumber\\
&&n^{\mu}=\frac{\rho\bar{\rho}}{2}[(r^{2}+a^{2}), -{\sf\Delta}, 0, a],\nonumber\\
&&m^{\mu}=-\frac{\bar{\rho}}{\sqrt{2}}[ia\sin\theta, 0, 1, \frac{i}{\sin\theta}],\nonumber\\
&&\bar{m}^{\mu}=-\frac{\rho}{\sqrt{2}}[-ia\sin\theta, 0, 1, -\frac{i}{\sin\theta}]; \nonumber\\
&&\kappa=\sigma=\nu=\lambda=\varepsilon=0,\nonumber\\
&&\rho=-\frac{1}{r-ia\cos\theta}, \quad \tau=-i\frac{\rho\bar{\rho}}{\sqrt{2}} a\sin\theta, \quad \mu=\frac{\rho^{2}\bar{\rho}}{2}{\sf\Delta}, \quad \gamma=\frac{\rho\bar{\rho}}{2}(r-M)+\mu, \nonumber\\
&&\pi=i\frac{\rho^{2}}{\sqrt{2}}a\sin\theta, \quad \beta=-\frac{\bar{\rho}}{2\sqrt{2}}\cot\theta, \quad \alpha=\pi-\bar{\beta}; \nonumber\\
&&\psi_{2}=\rho^{3}M; \nonumber\\
&&R=0;
\end{eqnarray}
where ${\sf\Delta}=r^{2}+a^{2}-2 M r$. The quantities in Eq. (A.6) are identical to the corresponding quantities given by Teukolsky in Ref. [3]

\textit{6. The variable-mass Kerr metric}
\begin{eqnarray}\label{A.7}
&&\kappa=\sigma=\lambda=0,\nonumber\\
&&\rho=\bar{\varrho}, \quad \tau=\frac{\bar{\varrho}^{2}}{\sqrt{2}}ia\sin\theta, \quad \varepsilon=\frac{1}{2}(\bar{\varrho}-\varrho), \quad \pi=-\frac{\varrho\bar{\varrho}}{\sqrt{2}}ia\sin\theta, \quad \nu=-\frac{\varrho^{2}\bar{\varrho}}{2}i\dot{M}ra\sin\theta, \nonumber\\
&&\mu=\frac{\varrho\bar{\varrho}^{2}}{2}(r^{2}+a^{2}-2Mr), \quad \gamma=-\frac{(\varrho\bar{\varrho})^{2}}{2}[M(-r^{2}+a^{2}\cos^{2}\theta)+ra^{2}\sin^{2}\theta], \nonumber\\
&&\beta=\frac{\bar{\varrho}}{2\sqrt{2}}(ia\varrho\sin\theta+ia\bar{\varrho}\sin\theta-\cot\theta), \quad \alpha=\frac{\varrho}{2\sqrt{2}}(2a^{2}\varrho\bar{\varrho}\sin\theta\cos\theta+\cot\theta),
\end{eqnarray}
where the dot denotes the derivative with respect to $v$.

\section{Some black hole solutions covered in three spacetime  families}

The physical interpretation of the parameters listed in the three tables of this appendix can be found in the corresponding references.

\begin{table}[htbp]
\centering
\begin{tabular}{l|r}
\hline
Black hole solution & Parameters \\
\hline
Schwarzschild & M \\
Schwarzschild-(A)dS & M, $\Lambda$ \\
Reissner-Nordstr$\ddot{o}$m & M, Q \\
Reissner-Nordstr$\ddot{o}$m-(A)dS & M, Q, $\Lambda$ \\
Gauss-Bonnet &  M, $\alpha$ \\
Gauss-Bonnet-(A)dS [66, 67] & M, $\Lambda$, $\alpha$ \\
Black holes in string-generated gravity models [68] & M, Q, $\alpha$ \\
Myers-Perry and its generalizations &  M, J \\
Dilaton  & M, Q, $\phi$ \\
Dilaton-Gauss-Bonnet &  M, $\phi$, $\alpha$ \\
Black universes  & M, $\phi$ \\
McVittie [69, 70] & M \\
Reissner-Nordstr$\ddot{o}$m metric in the FRW universe [71] & M, Q \\
Black holes in an expanding universe [72] & $Q_{T}$, $Q_{S}$, $Q_{S'}$, $Q_{S''}$ \\
Extremal magnetically charged black hole [73] & M, Q, $\phi_{0}$ \\
Supersymmetric RN-AdS [74] & M, Q, P, $\Lambda$ \\
Topological black hole [75] & M, $\Lambda$ \\
Stringy black holes [76] & $r_{0}$, $\delta_{2}$, $\delta_{5}$, $\delta_{6}$, $\delta_{p}$  \\
Garfinkle-Horowitz-Strominger [77] & M, Q, $\phi_{0}$ \\
Gibbons-Maeda dilaton [78] & M, Q, P \\
Born-Infeld [79] & M, Q, $\beta$ \\
Regular phantom black holes [80, 81] & M, b, c \\
Barriola-Vilenkin [82] & M, $\eta$ \\
Grumiller [83] & M, a, $\Lambda$ \\
Kiselev [84] & $r_{g}$, $r_{q}$ Q, $\Lambda$, $\omega_{q}$ \\
Quantum-corrected black holes [85] & M, $\alpha$ \\
\hline
\end{tabular}
\caption{General spherically symmetric spacetimes.\label{tab:1}}
\end{table}

\begin{table}[htbp]
\centering
\begin{tabular}{l|r}
\hline
Black hole solution & Parameters \\
\hline
Schwarzschild & M \\
Schwarzschild-(A)dS & M, $\Lambda$ \\
Reissner-Nordstr$\ddot{o}$m & M, Q \\
Reissner-Nordstr$\ddot{o}$m-(A)dS & M, Q, $\Lambda$ \\
Vaidya  &  \\
Vaidya-(A)dS & $\Lambda$ \\
Vaidya-Bonner [86] &  \\
Vaidya-Bonner-(A)dS [87] & $\Lambda$ \\
Radiating black holes with an internal global monopole [88] & $\eta_{0}$ \\
\hline
\end{tabular}
\caption{General Vaidya-type spacetimes.\label{tab:2}}
\end{table}

\begin{table}[htbp]
\centering
\begin{tabular}{l|r}
\hline
Black hole solution & Parameters \\
\hline
Pleba$\acute{n}$ski-Demia$\acute{n}$ski & m, e, g, a, $l$, $\alpha$, $\Lambda$, $\omega$ \\
Schwarzschild & m \\
Schwarzschild-(A)dS & m, $\Lambda$ \\
Reissner-Nordstr$\ddot{o}$m & m, e \\
Reissner-Nordstr$\ddot{o}$m-(A)dS & m, e, $\Lambda$ \\
Kerr  & m, a \\
Kerr-Newman & m, e, a \\
Kerr-Newman-(A)dS & m, e, a, $\Lambda$ \\
Taub-NUT & m, $l$ \\
Kerr-NUT & m, a, $l$ \\
Kerr-Newman-NUT & m, a, e, $l$ \\
Kerr-Newman-NUT-(A)dS & m, a, e, $l$, $\Lambda$ \\
C-metric  & m, $\alpha$ \\
Accelerated Kerr  & m, a, $\alpha$ \\
Accelerated Kerr-Newman & m, e, a, $\alpha$ \\
Accelerated Kerr-Newman-(A)dS & m, e, a, $\alpha$, $\Lambda$ \\
\hline
\end{tabular}
\caption{The complete family of black hole-like spacetimes.\label{tab:3}}
\end{table}

\section{The Bianchi identities in vacuum space-times}

The Bianchi identities for the vacuum case are given by [43]

\begin{eqnarray}\label{C.1}
&&D\psi_{1}-\bar{\delta}\psi_{0}=-3\kappa\psi_{2}+(2\varepsilon+4\rho)\psi_{1}-(4\alpha-\pi)\psi_{0}, \nonumber\\
&&D\psi_{2}-\bar{\delta}\psi_{1}=-2\kappa\psi_{3}+3\rho\psi_{2}-(2\alpha-2\pi)\psi_{1}-\lambda\psi_{0}, \nonumber\\
&&D\psi_{3}-\bar{\delta}\psi_{2}=-\kappa\psi_{4}-(2\varepsilon-2\rho)\psi_{3}+3\pi\psi_{2}-2\lambda\psi_{1}, \nonumber\\
&&D\psi_{4}-\bar{\delta}\psi_{3}=(\rho-4\varepsilon)\psi_{4}+(4\pi+2\alpha)\psi_{3}-3\lambda\psi_{2}, \nonumber\\
&&\Delta\psi_{0}-\delta\psi_{1}=(4\gamma-\mu)\psi_{0}-(4\tau+2\beta)\psi_{1}+3\sigma\psi_{2}, \nonumber\\
&&\Delta\psi_{1}-\delta\psi_{2}=\nu\psi_{0}+(2\gamma-2\mu)\psi_{1}-3\tau\psi_{2}+2\sigma\psi_{3}, \nonumber\\
&&\Delta\psi_{2}-\delta\psi_{3}=2\nu\psi_{1}-3\mu\psi_{2}+(2\beta-2\tau)\psi_{3}+\sigma\psi_{4}, \nonumber\\
&&\Delta\psi_{3}-\delta\psi_{4}=3\nu\psi_{2}-(2\gamma+4\mu)\psi_{3}+(4\beta-\tau)\psi_{4}.
\end{eqnarray}

\acknowledgments
This work was supported by the National Natural Science Foundation
of China under Grant No. 12175198.





\end{document}